\documentclass[sigconf]{acmart}
\AtBeginDocument{%
  \providecommand\BibTeX{{%
    \normalfont B\kern-0.5em{\scshape i\kern-0.25em b}\kern-0.8em\TeX}}}

\copyrightyear{2024}
\acmYear{2024}
\setcopyright{acmlicensed}\acmConference[CHI '24]{Proceedings of the CHI
Conference on Human Factors in Computing Systems}{May 11--16,
2024}{Honolulu, HI, USA}
\acmBooktitle{Proceedings of the CHI Conference on Human Factors in
Computing Systems (CHI '24), May 11--16, 2024, Honolulu, HI, USA}
\acmDOI{10.1145/3613904.3642257}
\acmISBN{979-8-4007-0330-0/24/05}

\acmConference[CHI '24]{Make sure to enter the correct
  conference title from your rights confirmation email}{May 11--16,
  2024}{Honolulu, HI, USA}

\usepackage{subcaption}
\usepackage{tabu}   % only used for the table example
\usepackage[export]{adjustbox}
\usepackage{xcolor}

\makeatletter
\renewcommand{\textcolor}[2]{\color{black}#2}
\makeatother

\begin{document}

%\title{Hearing Tumors: Interactive Shape Sonification for Breast Cancer Localization}
\title{Interactive Shape Sonification for Tumor Localization in Breast Cancer Surgery}
%\title[short title]{full title}


\author{Laura Sch\"{u}tz}
\orcid{0000-0002-5534-3903}
\affiliation{%
    % add CAMP chair to the affiliation and move Nassir to the back?
  \institution{Technical University of Munich}
  \streetaddress{Boltzmannstrasse}
  \city{Munich}
  \country{Germany}
  \postcode{85748}
}
\email{laura.schuetz@tum.de}

\author{Trishia El Chemaly}
\orcid{0000-0002-4234-3082}
\affiliation{%
  \institution{Stanford University}
  \streetaddress{Jane Stanford Way}
  \city{Stanford}
  \state{California}
  \country{USA}
}
\email{tchemaly@stanford.edu}

\author{Emmanuelle Weber}
\orcid{0000-0003-0277-394X}
\affiliation{%
  \institution{Stanford University}
  \streetaddress{Jane Stanford Way}
  \city{Stanford}
  \state{California}
  \country{USA}
}
\email{emmweber@stanford.edu}

\author{Anh Thien Doan}
\orcid{0009-0001-1680-3364}
\affiliation{%
  \institution{Stanford University}
  \streetaddress{Jane Stanford Way}
  \city{Stanford}
  \state{California}
  \country{USA}
}
\email{anhtdoan@stanford.edu}

\author{Jacqueline Tsai}
\orcid{0009-0009-7122-0557}
\affiliation{%
  \institution{Stanford University}
  \streetaddress{Jane Stanford Way}
  \city{Stanford}
  \state{California}
  \country{USA}
}
\email{jatsai@stanford.edu}

\author{Christoph Leuze}
\orcid{0000-0003-4564-8014}
\affiliation{%
  \institution{Stanford University}
  \streetaddress{Jane Stanford Way}
  \city{Stanford}
  \state{California}
  \country{USA}
}
\email{cleuze@stanford.edu}

\author{Bruce Daniel}
\orcid{0000-0003-0475-5892}
\affiliation{%
  \institution{Stanford University}
  \streetaddress{Jane Stanford Way}
  \city{Stanford}
  \state{California}
  \country{USA}
}
\email{bdaniel@stanford.edu}

\author{Nassir Navab}
\orcid{0000-0002-6032-5611}
\affiliation{%
    % add CAMP chair to the affiliation and move Nassir to the back?
  \institution{Technical University of Munich}
  \streetaddress{Boltzmannstrasse}
  \city{Munich}
  \country{Germany}
  \postcode{85748}
}
\email{nassir.navab@tum.de}

\renewcommand{\shortauthors}{Sch\"{u}tz et al.}

\begin{abstract}
    About 20 percent of patients undergoing breast-conserving surgery require reoperation due to cancerous tissue remaining inside the breast. Breast cancer localization systems utilize auditory feedback to convey the distance between a localization probe and a small marker (seed) implanted into the breast tumor prior to surgery. However, no information on the location of the tumor margin is provided. To reduce the reoperation rate by improving the usability and accuracy of the surgical task, we developed an auditory display using shape sonification to assist with tumor margin localization. Accuracy and usability of the interactive shape sonification were determined on models of the female breast in three user studies with both breast surgeons and non-clinical participants. The comparative studies showed a significant increase in usability (p<0.05) and localization accuracy (p<0.001) of the shape sonification over the auditory feedback currently used in surgery.
\end{abstract}




\begin{teaserfigure}
  \includegraphics[width=\textwidth]{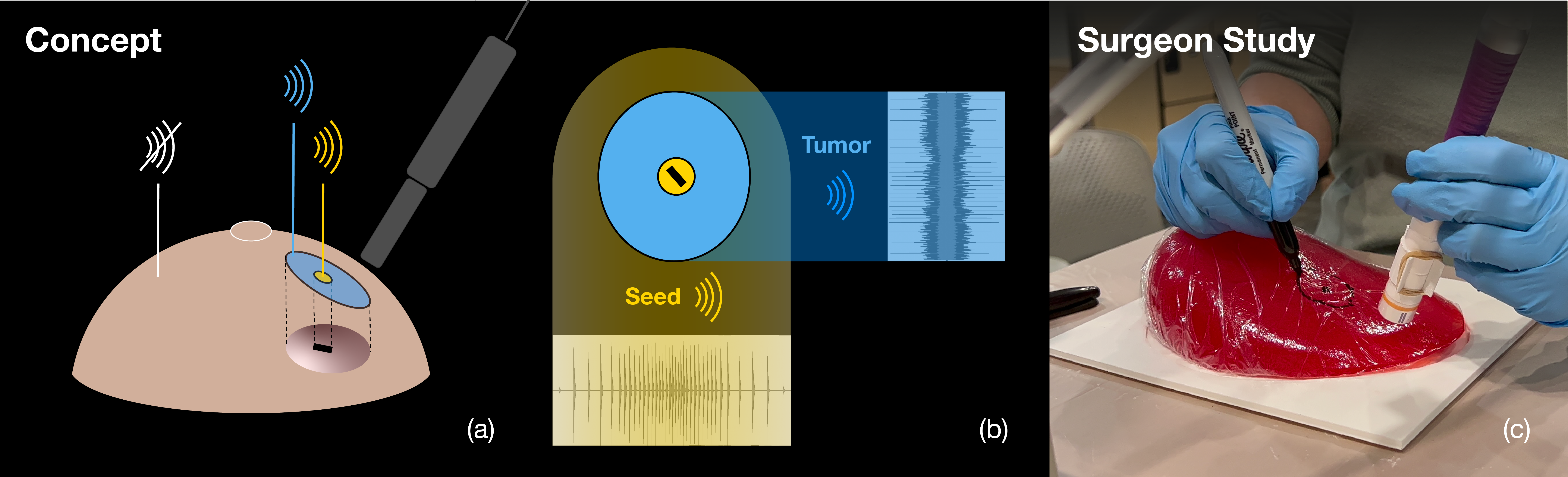}
  \caption{Left to right: (a) Shape sonification concept for breast tumor localization shown on a drawing of the female breast.
  (b) Top view: Yellow indicates the seed location. As the distance to the seed decreases, the frequency of the beating sound increases. Blue represents the tumor shape. A discrete synthesizer sound is triggered whenever the localization probe is above the tumor. 
  (c) A breast surgeon using shape sonification to localize a tumor on an agar breast model during Study 3.}
  \Description{Sub-figure (a) shows a three-dimensional view of the breast in a lying down position with an ellipsoid tumor and a seed, a small rectangular box, inside the breast. The tumor margin and seed location are projected onto the skin bottom up, and the two areas are marked on the skin with a blue ellipse for the tumor margin and a yellow ellipse for the seed area. Sub-figure (b) shows the same two ellipsoid shapes as well as the sound waves produced by the tumor and seed sound. The sound waveforms show the change in sound intensity over time. The seed sound waveform shows pikes that densify towards the middle of the seed. The tumor sound waveform shows a noisy wave that keeps the same sound intensity over the entire tumor area. Sub-figure (c) shows two hands with surgical gloves. One hand holds a black marker, the other a localization probe (a cylindrical handpiece). The person is in the process of drawing an ellipsoid shape on top of a red breast model wrapped in cling wrap.}
  \label{fig:teaser}
\end{teaserfigure}

\received{14 September 2023}
\received[revised]{26 November 2023}

\maketitle

\section{Introduction}

Breast cancer continues to be a significant health challenge, with its incidence surpassing that of any other cancer worldwide. In 2018 alone, over two million new cases of breast cancer were diagnosed, accounting for 23\% of all global cancer cases. The most common treatment for women diagnosed with breast cancer is a lumpectomy, a breast-conserving procedure that aims to remove the tumor and a small margin of surrounding healthy tissue \cite{bib1}. 
Male breast cancer is a rare condition, accounting for less than 1\% of all diagnosed breast cancer cases \cite{wehring2023pictorial,fentiman2006male}. Mastectomy has been the standard surgical approach for male breast cancer \cite{soni2023male}. However, recent studies have suggested that lumpectomy can be an oncologically safe treatment option for male patients as well \cite{de2019saving}.
Despite its effectiveness, a lumpectomy has limitations, including the potential for tumor-positive margins, i.e., cancerous tissue left inside the breast after the initial procedure. In this case, reoperation is necessary \cite{bib2}. Reported reoperation rates range from less than 10\% to more than 70\% \cite{bib2,bib3,bib4,bib5,bib6,bib7,bib8,bib12}. Another common challenge during lumpectomy is overexcision, where surgeons remove more tissue than necessary, leading to poorer cosmetic outcomes. While surgeons aim to remove the entire tumor with a 1-cm margin, resection volumes have been reported to be 2.3 to 2.5 times larger than the optimal resection volume \cite{bib13,bib14}.
\newline
Breast cancer localization systems have been developed to make excision more precise and thus decrease the rate of reoperations. Some localization systems, such as the Savi Scout®\footnote{Savi Scout®, Merit Medical Systems, Jordan, UT, USA} use sound feedback to convey the location of the tumor to the surgeon \cite{mango}. However, only the location of the seed, a small marker implanted into the tumor prior to surgery, is sonified. \textcolor{red}{Although the tumor size and location can be visualized using medical imaging, these images are obtained prior to surgery. Due to the deformability of breast tissue, the location and shape of the tumor will have changed from the position of the breast during preoperative imaging to the position in surgery.} This introduces a potential margin of error for the surgeon in determining the tumor's actual shape and margin location \textcolor{red}{at the time of surgery, even with the use of current localization systems}. \textcolor{red}{This work aims} to evaluate whether sonification of both the seed location and the tumor shape is effective in improving breast cancer localization accuracy and overall system usability. Novel auditory displays encoding both seed and margin location were evaluated (Figure~\ref{fig:teaser}). Beyond breast cancer surgery, these sonification strategies can be extended to various surgical tasks or even find application in multimodal user interfaces.
\newline
\newline
Our work provides the following main contributions:
\begin{enumerate}
\item We introduce shape sonification into surgical guidance systems.
\item We present an auditory display for breast cancer localization using multi-parameter sound mapping for simultaneous encoding of shape information (tumor margin) and point location (seed).
\item We report results from three user studies with four breast surgeons and 33 non-clinical participants comparing shape sonification to the current clinical sound feedback.
\item We provide evidence that shape sonification has the potential to improve the usability and accuracy of surgical localization tasks.
\end{enumerate}

\section{Related Work}
The work presented in this paper most closely relates to three areas of research, which the following section summarizes: auditory displays with a focus on shape sonification, recent experiments in sonification for surgical guidance, and breast cancer localization approaches.

\subsection{Shape Sonification}
Previous research in psychoacoustics and sensory substitution proves that the visual sensory channel can be substituted by the auditory channel for certain tasks. A study by Auvray et al. \textcolor{red}{[\citeyear{auvray}]} showed that objects can be recognized by their auditory representation and that auditory cues can be used for localization tasks. Meijer et al. \textcolor{red}{[\citeyear{meijer}]} conducted an experiment on auditory image representation. They converted images into corresponding sound patterns, proving that auditory representation can effectively retain visual image information.
Gerino et al. \textcolor{red}{[\citeyear{gerino}],} who evaluated multiple two-dimensional (2D) sonification techniques for shape recognition on touchscreens, were able to show that sonification is effective at conveying shape information to users. They introduced a shape sonification task which asked the user to explore an invisible 2D shape (e.g., a triangle or a \textcolor{red}{rectangle}) by moving their index finger along a line on a touch screen. For each position of the finger on this one-dimensional line, a "cross-section" of the shape was sonified and played to the user. 
Another study on geometric shape recognition by van den Doel et al. \textcolor{red}{[\citeyear{doel}]} focused on sound feedback for \textcolor{red}{blind people}.
Their system mapped images of basic geometric shapes to sound. During their study, the user explored the virtual image by moving a pointer over a graphics tablet (Wacom tablet) to explore the shape and was later asked to draw the perceived shape.

A tablet-based system was also used in a showcase by Javier Sanchez \textcolor{red}{[\citeyear{sanchez}],} enabling sound-based curve and shape recognition through pen and finger movement.
Tommasini et al. \textcolor{red}{[\citeyear{tommasini}]}, while also exploring invisible geometric shapes via sound feedback, were interested in quantifying the dynamic movements of a computer mouse during shape exploration. They showed differences in exploration strategies for the distinct geometric shapes.
A study evaluating shape recognition via gestures asked users to discriminate between concave and convex curves in three-dimensional (3D) space using sound \cite{boyer}. 
%The user was able to explore the 3D surface with hand gestures.

Our work introduces a new use case for shape sonification, namely breast cancer localization. The proposed system sonifies both the tumor shape and the location of a marker inside the tumor.

\subsection{Surgical Sonification}
Auditory displays are a promising means of guidance during medical procedures when visual focus is required on the surgical site, and a visual overlay in Augmented Reality (AR) might distract from the surgical task. Among works introducing sonification as an alternative or addition to visual surgical guidance systems are Matinfar et al. \textcolor{red}{[\citeyear{matinfar}]}, Roodaki et al. \textcolor{red}{[\citeyear{roodaki}]} and Schütz et al. \textcolor{red}{[\citeyear{schuetz}]}. Matinfar et al. \textcolor{red}{[\citeyear{matinfar}]} presented a four-dimensional sonification for surgical instrument alignment during pedicle screw placement. 
Schütz et al. \textcolor{red}{[\citeyear{schuetz}]} proposed an audiovisual guidance system for coil placement during transcranial magnetic stimulation by means of position and angle sonification. 
Roodaki et al. \textcolor{red}{[\citeyear{roodaki}]} evaluated the influence of sound feedback on the performance of medical precision tasks, such as needle placement in eye surgery. The audio guidance encoded location and angle information about the needle to the user. Their study showed increased angle alignment accuracy over a visual medical guidance system.
All three works showed that sonification is as effective as visualization in conveying tool location and angle information.
Two other studies presenting an auditory display for needle placement are Black et al. \textcolor{red}{[\citeyear{black}]} and Bork et al. \textcolor{red}{[\citeyear{bork}]}. Black et al. \textcolor{red}{[\citeyear{black}]} presented an auditory synthesis model using pitch and stereo panning parameter mapping for navigated needle placement, while Bork et al. \textcolor{red}{[\citeyear{bork}]} introduced an audiovisual AR system to improve occluded anatomy localization perception in 3D. The latter study showed enhanced needle placement accuracy when using auditory and visuotemporal guidance.
Another study evaluating the benefit of sonification in a surgical setting mapped the spatial position of a surgical tool tip with respect to a target location by encoding \textcolor{red}{the} direction and distance to said target position in the auditory display \cite{ziemer}. Directionality was represented via sound pitch and distance to \textcolor{red}{the} target via beat frequency. 

Unlike the studies presented in this section, which applied sonification approaches to surgical tool location and orientation, our work focuses on shape sonification. We present a novel sonic interaction with the tumor. Instead of sonifying the surgical tool alignment with some pre-planned position, we sonify the anatomical target.

\subsection{Breast Tumor Localization Techniques}
Current intraoperative localization methods of non-palpable breast lesions rely on medical imaging such as mammography, ultrasound, or magnetic resonance imaging (MRI) \cite{kasem2020savi}. Although the acquired imaging data is volumetric, scans are commonly presented \textcolor{red}{to the surgeon} as a series of 2D multiplanar images and thus demand experienced mental registration of these images to the patient to locate the tumor. Image-guided tumor localization is furthermore challenging as the patient's position during diagnostic imaging differs from their position during the surgical procedure. For instance, \textcolor{red}{a} breast MRI is typically acquired with the patient lying face down in a prone position, while a mammogram is taken with the patient standing and the breast compressed between two plates. In contrast, actual lumpectomy surgery is performed with the patients lying on their backs. 

To ensure precise localization, even after \textcolor{red}{the position is changed}, a trackable marker is embedded directly into the tumor, which can be localized regardless of breast tissue deformation. While wire-guided localization has been a common approach, its limitations prompted the development of a wireless alternative using radioactive seeds, which, however, requires substantial training for the safe handling of radioactive material \cite{kasem2020savi}. Several radiation-free and wireless methods, including the use of radiofrequency identification tags, magnetic seeds, and infrared reflectors, have emerged \textcolor{red}{\cite{ref12, ref13, kasem2020savi}}. These markers can be inserted at the time of pre-operative biopsy. 

An exemplary radiation-free localization system is the Savi Scout®. It involves a 12×1.6 mm electromagnetic wave
reflector (seed) activated by infrared light impulses generated by the console probe and two antennas that allow the reflection of the electromagnetic wave signal back to the probe \cite{ref16}. The system provides real-time proximity information of the detection probe to the seed through auditory feedback \cite{kasem2020savi}. A recent analysis of over 800 cases revealed a significant decrease in reoperation to 12.9\% when using the Savi Scout® \cite{kasem2020savi}.

Although marker-based localization systems can help determine the location of the seed within the tumor, they do not provide information about the tumor margin location or its overall shape. Inferring these details from preoperative medical images can be a cognitively demanding and error-prone task. To address this challenge, several studies have explored the use of AR to assist with breast tumor localization \cite{perkins, perkins2, gouveia, lan}. Perkins et al. \textcolor{red}{[\citeyear{perkins}]} demonstrated a Mixed Reality surgical planning system for breast tumor targeting in seven patients. The location of the tumor was marked on the skin of the breast by tracing the virtual tumor shape. In a preliminary 2D perceptual task, they evaluated the tracking accuracy of the localization system. They reported an overlap of the drawn and the virtual shape using the Dice coefficient (ranging from 0.56 to 0.95). For the patient study, they reported a Dice coefficient of about 0.2 comparing the shape drawn using a HoloLens and the ground truth shape \cite{perkins2}.
Unlike Perkins et al. \textcolor{red}{[\citeyear{perkins}]}, who introduced a planning system to be used before incision, Gouveia et al. \textcolor{red}{[\citeyear{gouveia}]} performed a live lumpectomy using AR guidance.
Lan et al. \textcolor{red}{[\citeyear{lan}]} proposed the use of a fiber optoacoustic guide in combination with a tablet-based AR interface for lumpectomy. Their study showed a localization accuracy of 0.25mm of the tip of the fiber optoacoustic guide. However, no information about the tumor shape was presented to the user. Another recent work in the field of breast tumor localization introduced an AR visualization system for ultrasound breast biopsy, which offers segmented lesion visualization \cite{costa2023augmented}. Their system demonstrated improved accuracy, as evidenced by the distance from the needle tip to the lesion being reduced to 5.09 mm.

Unlike the above studies presenting visual augmentation approaches, we propose to add to the state-of-the-art practices in breast tumor localization by providing an auditory display for seed-based localization systems. We \textcolor{red}{aim} to enhance tumor excision precision by providing information about the tumor margin location in addition to the seed location sonified in the state-of-the-art solutions.

\begin{figure*} [t!]
    \centering\includegraphics[width=\textwidth]{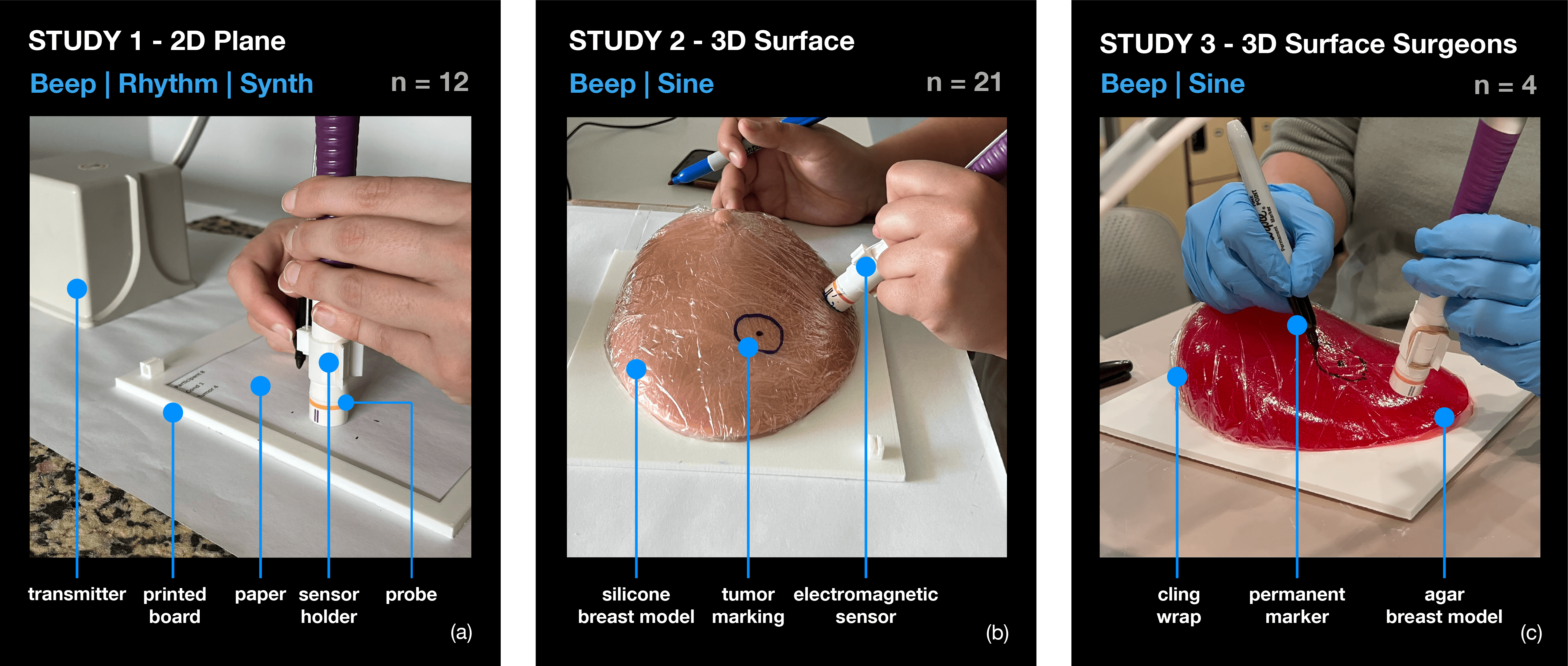}
    \caption{Experiment setup and sonifications used in the three user studies. The setup included the transmitter of the trakSTAR™ electromagnetic (EM) tracking system, the Savi Scout® radar localization probe tracked using an EM sensor, a three-dimensional (3D) printed holder for the EM sensor, and a pen. Study 1 used a 3D printed board \textcolor{red}{to hold} the paper in place. Study 2 and 3 used a 3D printed board for holding the breast models in place.
    }
    \Description{The three experiment setups are shown in order of Study 1 to Study 3 from left to right. Each picture shows a participant holding the probe in one hand and pen in the other to locate and mark a tumor margin. Study 1 uses a white paper sheet for marking, Study 2 a beige silicone breast, and Study 3 a red gelatinous breast model.
    }
    \label{fig:3_studies}
\end{figure*}

\section{Experiments}

% hypothesis / questions we wanted to answer with the studies
We conducted three user studies to investigate whether shape sonification can increase the usability and accuracy of the tumor localization task in lumpectomy. Repeat testing helped us to incrementally refine the shape sonification design. For the first study, the use context was abstracted to a 2D plane (Figure~\ref{fig:3_studies}(a)). The second study evaluated a refined shape sonification design on a silicone model of the female breast (Figure~\ref{fig:3_studies}(b)), and the third study tested the final sonification design on 16 agar breast models with breast surgeons (Figure~\ref{fig:3_studies}(c)).
All studies were approved by the university's Institutional Review Board (IRB) office. 
In particular, we sought to test the following hypotheses: 
\newline\newline
\textbf{Hypothesis 1 (H1)}: Shape sonification (Rhythm, Synth, Sine) significantly increases the localization \textbf{accuracy} compared to the current clinical auditory feedback (Beep).
\newline\newline
\textbf{Hypothesis 2 (H2)}: Shape sonification (Rhythm, Synth, Sine) significantly enhances the \textbf{usability} of the auditory display compared to the current clinical auditory feedback (Beep).
\newline\newline
\textbf{Hypothesis 3 (H3)}: Shape sonification (Sine) significantly decreases the amount of \textbf{excess} healthy breast \textbf{tissue resected} compared to the current auditory feedback (Beep).
%\item H4: Breast surgeons perform better (localization accuracy and usability) than non-surgeons.

%%%%%%%%%%%%%%% 1 1 1 1 1 1 1 1 1 1 1 1 1 %%%%%%%%%%%%%%%%%%%%

\section{Study 1 - 2D Plane}

% short summary of the objectives of study 1 and why we did it
We conducted a repeated-measures within-subject study to compare two shape sonification designs to the current clinical auditory feedback.

\subsection{Participants}
Twelve volunteers took part in the study, five women and seven men. The participants had an average age of 31 years, with a standard deviation of 9.74 years. 
The study included three postdoctoral researchers, one radiologist, three medical students, three bioengineering students, \textcolor{red}{and} two students in product design. None of the participants reported having any hearing impairments \textcolor{red}{or prior experience in using interactive sonification for shape localization}.

\begin{figure*} [t!]
    \centering\includegraphics[width=\textwidth]{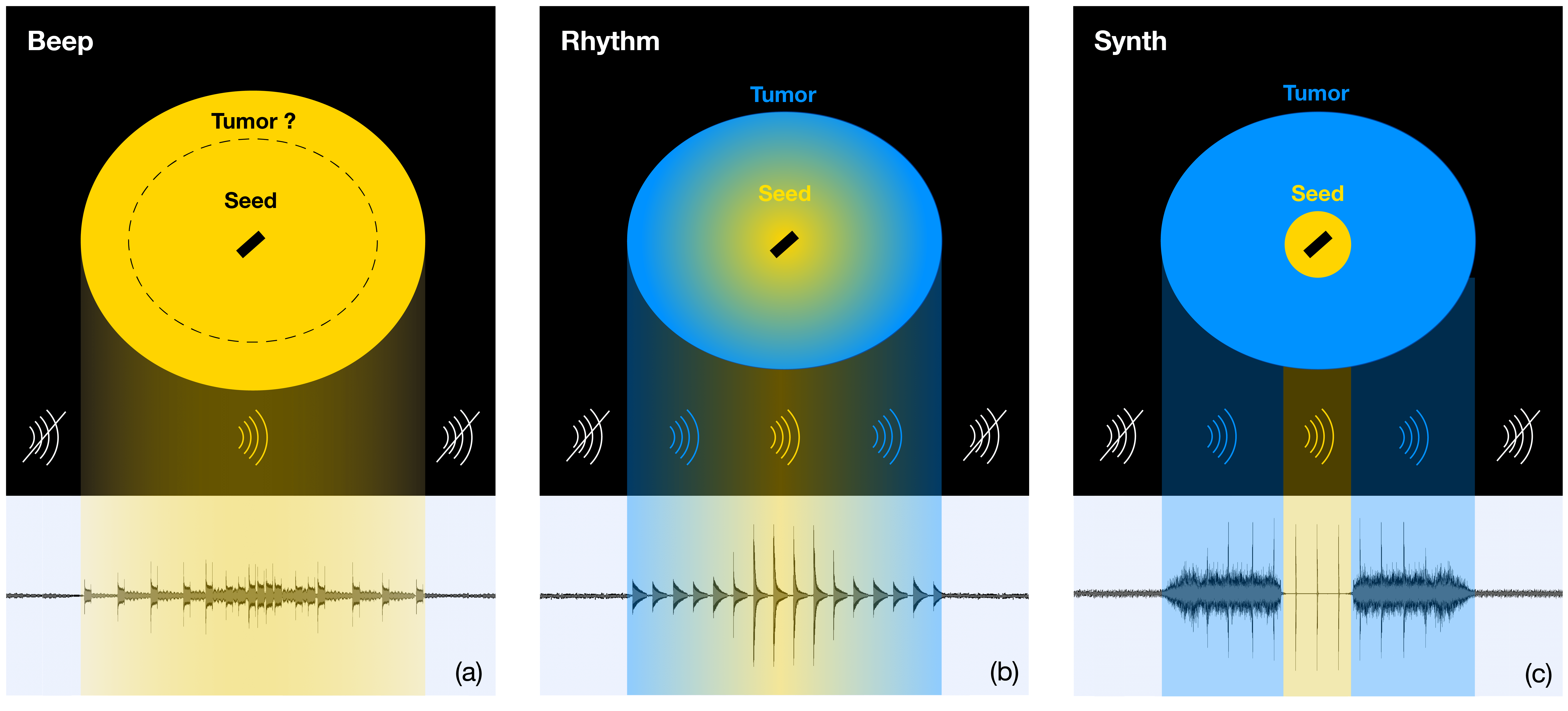}
    \caption{Sonification models in Study 1. \textcolor{red}{Sounds indicating the seed location are colored in yellow, sounds indicating the tumor shape are colored in blue}: (a) Beep (clinical status quo) and the newly proposed (b) Rhythm and (c) Synth shape sonifications are compared. (a) Beep uses a beeping sound \textcolor{red}{(yellow)} varying in frequency. (b) Rhythm uses one beating sound: the seed is represented by a xylophone \textcolor{red}{(yellow)}, the tumor by a marimba \textcolor{red}{(blue)}. The Rhythm sonification interpolates between the two instruments for the area between the tumor margin and the seed. (c) Synth uses two sounds to distinguish between margin and seed location: a continuous synthesizer sound \textcolor{red}{(blue)} represents the tumor margin, and a ticking sound \textcolor{red}{(yellow)} represents the seed. No interpolation is applied. The sounds are distinct.}
    \Description{The figure shows three simplified illustrations of the tumor shape and seed. Overlaid atop these illustrations are yellow and blue areas indicating the spatial localization of the sonifications and their sounds' waveform.}
    \label{fig:sonifications_s1}
\end{figure*}

\subsection{Apparatus}
\subsubsection{Hardware and Software Setup}
To enable the sonification study, we had to set up a surgical localization system. The system consisted of a Savi Scout® radar localization probe, an electromagnetic (EM) tracking system (the trakSTAR™ EM 6 degrees-of-freedom tracking solution)\footnote{trakSTAR™, NDI, Waterloo, Ontario, Canada} with a mid-range transmitter, and a graphics processing unit accelerated laptop for running the sonification and tracking software. The transmitter of the EM tracking system was placed on a table and faced the area where the user performed the instructed task (Figure~\ref{fig:3_studies}(a)). 

The EM tracking system was connected to the laptop via USB, and the open-source library Plus Toolkit\footnote{Plus Toolkit (https://plustoolkit.github.io/)} was used for real-time streaming of the position tracking and sensor data to Unity\footnote{Unity (https://unity.com/)} \textcolor{red}{(Long Term Support Release 2020.3.21f1)}. In Unity, the distance between the radar localization probe and a virtual 3D model of a tumor was computed using the \emph{closestPoint()} method, which returns the closest point on the surface of the tumor object from the probe tooltip as a point in 3D space. The distance between this closest point on the surface of the tumor and the tip of the probe, as well as the distance between the probe and the virtual seed, were streamed to Wekinator\footnote{Wekinator (http://www.wekinator.org/)}, an interactive machine learning system for music composition \cite{fiebrink}. Wekinator used a Neural Network to map the distance measures from Unity to the output parameters used for sound synthesis in ChucK\footnote{ChucK (https://chuck.stanford.edu/)}, an audio programming language for real-time sound synthesis \cite{wang}. Unity, Wekinator, and ChucK communicated via Open Sound Control (OSC), a network protocol for interactive computer music \cite{wright}.

\subsubsection{Tracking and Registration}
While the Savi Scout® system provides audible feedback on the proximity of the probe to the implanted seed, it does not provide numerical information on the position and orientation required for our study. Therefore, we required a tracking method providing real-time position and orientation of the probe to train the sonification model \textcolor{red}{accurately}. Clinical navigation systems often use external tracking hardware. Besides optical tracking solutions, EM tracking is a popular alternative. EM tracking detects the position and orientation of a wired sensor inside a magnetic field created by a field transmitter. While the accuracy of EM tracking is susceptible to the presence of ferrous metals and conductive materials that may distort the magnetic field, we chose EM tracking over optical tracking to avoid line-of sight issues\cite{figueras2014surgical}. 

The Savi Scout® probe was tracked using a 0.9 mm diameter, six degrees-of-freedom sensor at the tip of a shielded flexible cable. We fixed the position of the EM sensor on the probe using a 3D printed holder. We also 3D printed a custom board to hold a 15x15 cm white drawing paper in place. The OBJ files of the 3D prints were imported into Unity for alignment.

We performed a rigid registration of the virtual board to the 3D printed physical board. First, four fiducial markers were placed on the physical board by tapping the corners with the tracked probe. Then, four corresponding virtual fiducials were placed onto the corners of the virtual board in Unity to achieve the rigid registration. The physical board was fixed to a table, allowing the setup to stay steady during the experiment.

\subsection{Sonification}
The distance to sound mapping was done in Wekinator using a multilayer perceptron Neural Network with two hidden layers per input parameter. The model was trained on 80 to 100 recordings of the probe position for each sonification model and the manually set desired sound parameter values. For the purpose of this study, three sonification models were created: one mimicking the clinical status quo (Beep) and two newly proposed shape sonifications (Rhythm, Synth).

\subsubsection{Beep (status quo)} The distance between the tip of the probe and the location of a virtual seed (point) inside the tumor object was used as input to the Neural Network. Two output parameters were mapped to \textcolor{red}{the} volume and frequency of a beeping sound (Figure~\ref{fig:sonifications_s1} (a)). The baseline beep sonification used a sine wave oscillator at 440 Hertz (Hz). As the distance between the probe and the tumor decreased, the frequency of the beeps increased. This condition imitated the current sound feedback used by the Savi Scout® system. No information about the tumor shape or margin location was included in this condition.

\subsubsection{Rhythm} The first of the two proposed shape sonification models used both the distance to the tumor margin and the distance to the seed as input parameters to convey both the shape and point information within the same beating sound (Figure~\ref{fig:sonifications_s1} (b)). Rhythm used a constant beat with \textcolor{red}{a} change in instrument to indicate the tumor margin and seed location to the user. The sound was synthesized using the \emph{ModalBar} instrument in ChucK. The pitch was set to 330 Hz. When the probe was positioned above the tumor border, our system synthesized a distinct marimba beat. When the probe was above the seed location, a higher-pitched and clearer xylophone beat was played. For any location inside the tumor between the border and the inside seed location, the two sounds of the marimba and xylophone beat were interpolated. Four output parameters from Wekinator were mapped to \textcolor{red}{the} volume and timbre of the two beat sounds.

\subsubsection{Synth} The second shape sonification used the same input parameters as Rhythm to create two distinct sounds: a continuous sound to represent the shape and a beat to represent the seed location (Figure~\ref{fig:sonifications_s1} (c)).
This sonification separated the areas of interest into discrete zones. The Synth sonification model played a continuous synthesizer sound \emph{(musical note C4)} when the probe was above the tumor area and no sound when the probe was outside the area. For the seed location, a ticking beat sound (ChucK \emph{ModalBar} at 660 Hz) \textcolor{red}{was} synthesized to indicate the point of interest. Three output parameters from the Neural Network were mapped to \textcolor{red}{the} volume and timbre of the ticking seed sound and \textcolor{red}{the} volume of the continuous tumor sound.

\subsection{Procedure}
The three sonifications (Beep, Rhythm, Synth) were presented to each participant in a random order. We began the study session
by administering a pre-task questionnaire, including questions on demographic background, medical expertise, and video game experience. After the pre-task questionnaire, the tutorial phase began. The localization task and respective sonification concept \textcolor{red}{were} explained to the participant, followed by a trial run, which was not time-constrained. The tutorial phase was followed by the testing phase. The volunteer was tasked to mark the tumor margin and seed location on a paper sheet based on the sound feedback. The paper was exchanged after each tumor. Since the Beep condition lacked information about the tumor shape, the participant received a piece of paper with a picture of the tumor and seed showing the seed location relative to the tumor.
For each sonification, the participant was asked to localize six shapes randomly selected from a pool of 15. Once six tumors were located, a post-task questionnaire was filled out by the participant. The questionnaire consisted of a raw NASA-TLX (Task Load Index) questionnaire \cite{hart} to evaluate the participants' perceived task load, a System Usability Scale (SUS) \cite{lewis} to assess the user-perceived usability of the sonification, and several study-specific questions aimed at collecting qualitative feedback regarding their subjective impressions of the sonifications. This process was then reiterated twice more for the remaining sonifications.

\begin{figure*}
    \centering
    \subfloat[Sørensen-Dice coefficient (0-1; larger better)]
    {\resizebox*{5.75cm}{!}{\includegraphics{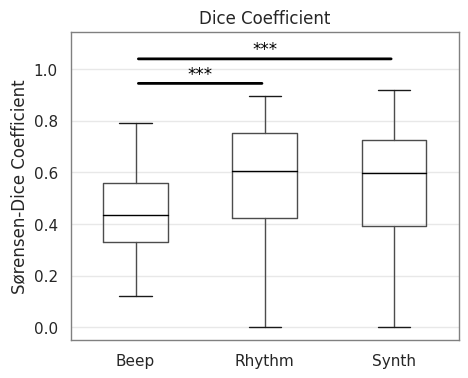}}}\hspace{6pt}
    \subfloat[Area ratio (1 equals perfect alignment of ground truth and drawn tumor shape)]
    {\resizebox*{5.58cm}{!}{\includegraphics{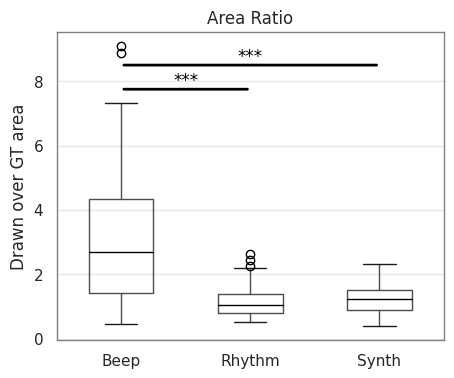}}} \hspace{6pt}
    \subfloat[Intercentroid distance (between ground truth and drawn seed location)]
    {\resizebox*{5.75cm}{!}{\includegraphics{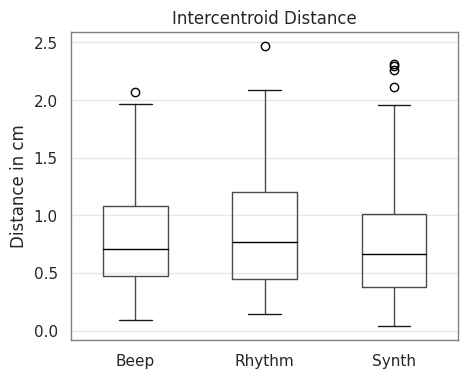}}}
    \caption{Box plots on localization accuracy in Study 1\textcolor{red}{. *** = p < 0.001}}
    \Description{Box plots of Sørensen-Dice coefficient (a), area ratio (b) and intercentroid distance (c) for the three sonification conditions Beep, Rhythm and Synth evaluated in Study 1.}
    \label{fig:boxplots_s1}
\end{figure*}

\subsection{Data Analysis}
\begin{figure}[hb]% specify a combination of t, b, p, or h for top, bottom, on its own page, or here
  \centering % avoid the use of \begin{center}...\end{center} and use \centering instead (more compact)
  \includegraphics[width=\linewidth]{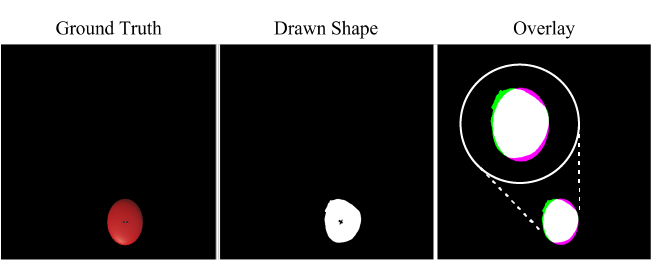}
  \caption{%
  Display of the segmented ground truth, drawn shape, and their overlay from one of the user experiments guided by the Synth sonification model. In the overlay image, the ground truth appears in white and pink, while the drawn shape appears in white and green. For the above case, the Sørensen–Dice coefficient was 0.92.%
  }
  \Description{Three black rectangles are displayed next to each other. The left-most includes a red tumor object, the center one a white surface with rough edges, and the right one an overlay of the two shapes and the resulting overlaps.}
  \label{fig:dice}
\end{figure}

To evaluate the accuracy of the shape sonification, we captured images of the 15 different tumor shapes in Unity in 2D view. The shapes' size, orientation, and location with respect to the drawing paper were visible on the images. The images were resized to 15x15 cm to match the size of the paper. In a next step, we used Matlab to segment the images of the virtual tumor shapes to obtain a ground truth (Figure~\ref{fig:dice}). By applying a threshold to the red color channel, we obtained the segmented shape of the ground truth tumors. Similarly, the corresponding ground truth for the seed locations \textcolor{red}{was} extracted from the Unity images.
%by summing the values of the three color channels (red, green, and blue) and checking if the sum is below a certain threshold.

The 15x15 cm sheets of paper used by the participants to draw the tumor shapes were scanned and later segmented in Matlab. Segmentation of the drawn tumors and corresponding seeds was achieved by finding connected components in the binary images using 8-connected neighborhoods. The tumor in each image was isolated as the largest circular connected component, while the corresponding seed was isolated as the largest connected component inside the drawn tumor area (Figure~\ref{fig:dice}). To isolate the drawn tumor, the circularity component was calculated using the following equation:

\begin{equation}
Circularity = \left(\frac{4\pi A}{P^2}\right)\left(1 - \frac{0.5}{r}\right)^2 \quad where \quad r = \frac{P}{2\pi} + 0.5
\end{equation}

where $A$ represents the area of the object, $P$ represents the perimeter, and $r$ is a parameter calculated from the perimeter.

The Sørensen–Dice coefficient, area ratio, and intercentroid distance were calculated to compare the drawn shape (DS) to the ground truth (GT) using the following equations \cite{sokal}:

\begin{equation}
\textnormal{Sørensen-Dice} = \frac{2|DS \cap GT|}{|DS| + |GT|}
\end{equation}

\begin{equation}
Area\ Ratio = \frac{Area_{DS}}{Area_{GT}}
\end{equation}

\begin{equation}
Intercentroid\ Distance = \sqrt{(x_{DS}-x_{GT})^2 + (y_{DS}-y_{GT})^2}
\end{equation}

\subsection{Results}
\subsubsection{Accuracy}
The accuracy data measured using the Sørensen Dice coefficient (1), area ratio (2), and intercentroid distance (3) were not normally distributed. \textcolor{red}{Out of the 72 datapoints (six repetitions per participant) twelve} outliers were removed using the interquartile range method. Since the study followed a within-subject design, a Friedman test was used to determine significance between conditions for the three measures of accuracy. The Friedman test showed a significant difference between conditions for the Dice coefficient (p$<$0.001) and the area ratio (p$<$0.001). No significant difference in intercentroid distance between sonifications was reported (Figure~\ref{fig:boxplots_s1}(c)). A paired Wilcoxon test with Holm correction was chosen to compare between conditions. The pairwise tests showed a significantly increased Dice coefficient (p$<$0.001) for both proposed shape sonifications (Rhythm and Synth) over the point sonification (Beep) (Figure~\ref{fig:boxplots_s1}(a)). Likewise, a significantly smaller area ratio (p$<$0.001) was recorded for the novel shape sonifications over Beep (Figure~\ref{fig:boxplots_s1}(b)). The significant increase in Dice coefficient proves that the use of the shape sonifications results in a larger overlap of the ground truth tumor shape and the drawn tumor shape, indicating an increased localization accuracy when using shape sonification. Means and standard deviations of all accuracy measures are summarized in Table~\ref{tab:means_s1}.

\subsubsection{Usability}
The responses to the NASA-TLX questionnaire on subjective workload were normally distributed. \textcolor{red}{No values were }outside the mean $\pm$ \textcolor{red}{three} std\textcolor{red}{, so no outliers} were excluded from the sample. A repeated measures ANOVA showed no significant difference between conditions \textcolor{red}{for the overall task load}. \textcolor{red}{Out of the NASA-TLX subscales, a significant difference was found for the performance scale. Paired samples t-tests showed a significant reduction (p$<$0.05) in \textcolor{red}{the} subjectively-perceived performance of Rhythm and Synth over Beep.}
A Shapiro-Wilk test showed that the SUS questionnaire data was not normally distributed. \textcolor{red}{T}he interquartile range method \textcolor{red}{was used to identify and remove one outlier}. A Friedman test indicated a significant difference between conditions (p$<$0.001). A post-hoc paired Wilcoxon test with Holm correction was used to compare sonifications. The pairwise Wilcoxon test showed that the Synth sonification's usability was perceived as significantly better than Beep (p$<$0.001) and Rhytm (p$<$0.05). A comparison of Rhythm and Beep did not yield significant differences in usability scores.
The Beep condition received the worst usability rating. Table~\ref{tab:means_s1} lists means and standard deviations of the NASA-TLX and SUS scores for all sonifications.

\begin{table} [pb]
    \caption{Means and standard deviations (std) of localization accuracy and sonification usability for all conditions in Study 1: Sørensen–Dice coefficient ([0,1]; larger better), area ratio (closer to 1 better), intercentroid distance (in cm; smaller better), \textcolor{red}{overall} task load\textcolor{red}{, individual NASA-TLX subscales }([0, 100]; smaller better) and usability (SUS, [0, 100]; larger better)}
    \label{tab:means_s1}
    \begin{adjustbox}{max width=\linewidth}
    \begin{tabular}{lccc}
        \toprule
        n = 12  & Beep & Rhythm & Synth \\
        \midrule
        Sørensen–Dice Coefficient & 0.44 $\pm$ 0.17 & 0.57 $\pm$ 0.23 & 0.57 $\pm$ 0.21 \\
        Area Ratio & 3.04 $\pm$ 1.99 & 1.15 $\pm$ 0.49 & 1.23 $\pm$ 0.46 \\
        Intercentroid Distance (cm) & 0.79 $\pm$ 0.41 & 0.89 $\pm$ 0.54 & 0.81 $\pm$ 0.59 \\
        NASA-TLX \textcolor{red}{- Overall} & 36.26 $\pm$ 10.36 & 42.03 $\pm$ 12.16 & 35.81 $\pm$ 17.02 \\
        \textcolor{red}{NASA-TLX - Mental Demand & \textcolor{red}{51.25 $\pm$ 29.63} & \textcolor{red}{39.28 $\pm$ 18.28} & \textcolor{red}{35.45 $\pm$ 17.85}} \\
        \textcolor{red}{NASA-TLX - Physical Demand & \textcolor{red}{18.00 $\pm$ 14.95} & \textcolor{red}{25.38 $\pm$ 19.75} & \textcolor{red}{21.87 $\pm$ 17.19}} \\
        \textcolor{red}{NASA-TLX - Temporal Demand & \textcolor{red}{33.18 $\pm$ 24.33} & \textcolor{red}{30.83 $\pm$ 20.36} & \textcolor{red}{15.00 $\pm$ 9.35}} \\
        \textcolor{red}{NASA-TLX - Performance & \textcolor{red}{32.92 $\pm$ 21.47} & \textcolor{red}{55.36 $\pm$ 21.52} & \textcolor{red}{65.00 $\pm$ 25.66}} \\
        \textcolor{red}{NASA-TLX - Effort & \textcolor{red}{37.50 $\pm$ 22.61} & \textcolor{red}{53.93 $\pm$ 19.82} & \textcolor{red}{43.64 $\pm$ 21.02}} \\
        \textcolor{red}{NASA-TLX - Frustration & \textcolor{red}{49.17 $\pm$ 31.32} & \textcolor{red}{46.79 $\pm$ 23.91} & \textcolor{red}{28.64 $\pm$ 26.62}} \\
        System Usability Scale & 57.24 $\pm$ 13.85 & 65.00 $\pm$ 19.66 & 85.47 $\pm$ 10.15 \\
        \bottomrule
    \end{tabular}
    \end{adjustbox}
\end{table}

\begin{figure*}
    \includegraphics[width=0.9\textwidth]{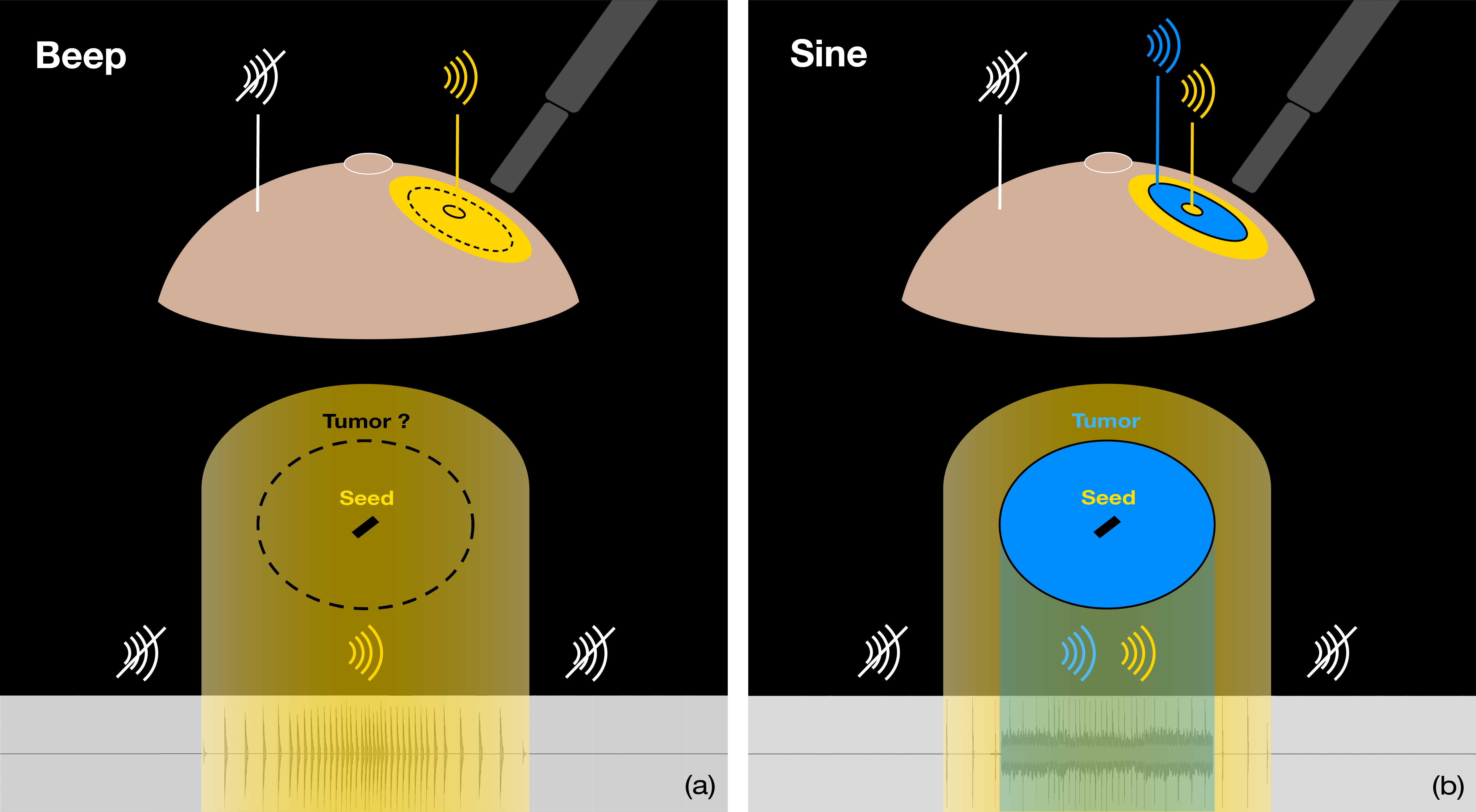}
    \caption{Sonification models in Study 2 and 3; \textcolor{red}{sounds indicating the seed location are colored in yellow, the sound indicating the tumor shape is colored in blue}: (a) Beep (clinical status quo) and (b) Sine (new shape sonification)}
    \Description{Sub-figure (a) and (b) show a three-dimensional view of the breast in a lying down position paired with a schematic drawing of the tumor and the seed. Sub-figure (a) shows a yellow area over the entire tumor shape. A sound wave image associated with the yellow area is the sound used in the Beep sonification. Sub-figure (b) shows a yellow ring around the tumor and a yellow area around the seed inside the tumor. The remainder of the tumor is colored blue. The yellow and blue areas are again associated with their respective waveform images.}
    \label{fig:sonifications_s2}
\end{figure*}

\subsubsection{Qualitative Feedback}
We asked the participants to rank the three sonifications. 75\% of the participants ranked Synth first. 58\% of the participants ranked them in this order: Synth, Rhythm, Beep. They elaborated that the continuous sound representing the tumor shape and the ticking sound representing the seed were more easily distinguishable in the Synth than the Rhythm sonification. To quote three participants: \textit{"I really like the consistent [Synth] sound, it made the boundary easier to find."}, \textit{"The [Rhythm] sounds were very pleasant, but I got confused about which tone indicated the seed."}, \textit{"[Rhythm] was hard to use to trace the outline."}. To improve the most preferred Synth sonification, three participants suggested adding a \textit{"distinct sound in the exact location of the seed"} to increase the localization accuracy.
The Beep sonification was perceived as \textit{"fast and intuitive"}, however, the sonification provided no \textit{"tumor shape information nor a direction to seed from [the] current position"}.

\subsection{Discussion}
Our results show that the use of shape sonification leads to a significant increase in localization accuracy. The Sørensen–Dice coefficient reports a greater overlap of the drawn tumor and the ground truth tumor for both novel shape sonifications compared to the clinical standard simulated by the Beep sonification. 
Besides measuring shape overlap, we also evaluated \textcolor{red}{the} area ratio. We were able to report a substantially decreased area ratio of drawn shapes over ground truth shapes for both shape sonifications compared to the status quo sonification.
Due to breast surgeons describing their mental model of the tumor location at the time of initial localization as that of a 2D shape on the surface of the breast, the preliminary sonification models were initially tested on a 2D plane. However, this experiment setup deviates from the real scenario, where the system is used on the curved surface of the breast.

%%%%%%%%%%%%%%% 2 2 2 2 2 2 2 2 2 2 2 %%%%%%%%%%%%%%%%%%%%

\begin{figure*} [h!]
    \centering
    \includegraphics[width=\textwidth]{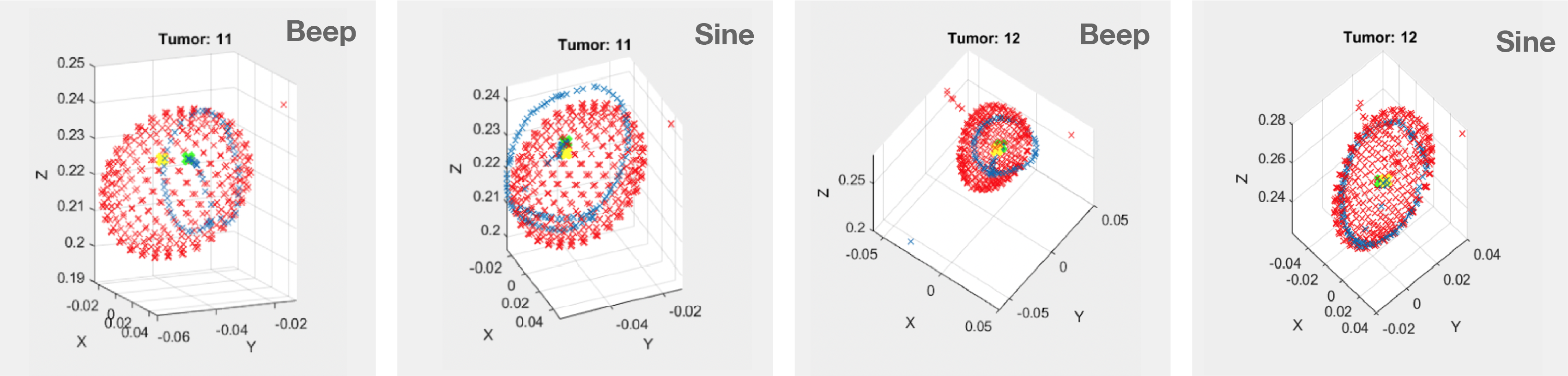}
    \caption{Matlab plots of the marking recordings of the ground truth tumor shape (red) and seed location (yellow), as well as the participant-drawn tumor margin (blue) and seed location (green) for two exemplary tumors (11, 12) using Beep and Sine}
    \Description{Four plots of location recordings represented as small crosses located in a 3D coordinate system are shown. The drawn marking using the Beep sonification is not as aligned with the ground truth tumor as the Sine sonification marking. The ground truth and drawn seed locations are located close to each other for all four plots.}
    \label{fig:matlab_plots}
\end{figure*}

\section{Study 2 - 3D Surface}

A second study was conducted to evaluate the auditory displays on a model of the breast. By using a breast phantom, the experimental setup simulated the surgical scene, and the results are more applicable to clinical research.

\subsection{Participants}
A total of 21 volunteers with a mean age of 26\textcolor{red}{.}5 $\pm$ 3\textcolor{red}{.}4 years participated in the study. \textcolor{red}{Twelve} indicated to be men and 9 to be women. 
The study included a radiologist, three postdoctoral researchers in Radiology, one medical student, nine graduate students in Bioengineering, four students studying Management, and three students specializing in Product Design. None of the participants reported to have any hearing impairments.
\textcolor{red}{Four participants indicated prior experience in using the interactive shape sonification.}
%\textcolor{red}{Postdoctoral researchers, PhD, graduate, and undergraduate students from the disciplines of computer science, design, biomedical engineering, business, and medicine were among the participants.}

\subsection{Apparatus}
This study used the same hardware and software setup as Study 1 except for the use of a silicone breast model instead of a paper sheet for task execution (Figure~\ref{fig:3_studies}(b)). 
For fabrication of the silicone breast phantom, we first poured a mold using casting silicone by Pixiss\footnote{Pixiss, Grand Rapids, MI, United States}. The 3B Scientific\footnote{American 3B Scientific, Tucker, GA, United States} Breast Self Examination Model 1 acted as the male part in the mold-making process. After the mold had cured for 12 hours, we used Smooth-On\footnote{Smooth-On, Macungie, PA, United States} Ecoflex™ 00-20, a "stretchy" silicone mixed with beige silicone pigment to pour the breast phantom. Silicone was chosen to replicate the physical properties of breast tissue, namely deformability and flexibility. A digital 3D model of the breast was segmented from an MRI scan of the breast phantom. The OBJ file was imported into Unity for registration.

In order to facilitate the marking of the tumor shape on the model's surface, we covered the phantom with cling wrap. After marking two to three tumors, we replaced the cling wrap to prevent overlapping markings. To secure the breast phantom in place, we used a custom 3D-printed board with a recess designed to match the breast model's shape. Both the OBJ files of the board and the breast were imported into Unity. The same rigid registration process as in Study 1 was performed to align the virtual and real experiment setups.

\subsection{Sonification}

The sonifications were implemented the same way as in Study 1. For this study, the status quo auditory feedback (Beep) was compared to one new shape sonification (Sine).

\subsubsection{Beep (status quo)}
Again, only the distance between probe and seed was sonified by Beep (Figure~\ref{fig:sonifications_s2}(a)). The sonification model from Study 1 was slightly changed to enhance its similarity to the current clinical sound feedback. Instead of the sine wave oscillator, the ChucK ModalBar instrument was used at 200 Hz. In addition, the frequency of the beat was increased. Three output parameters from Wekinator were mapped to sound volume and beat frequency.

\subsubsection{Sine}
Once more, the distance from the probe to the tumor margin and the seed was mapped to sound parameters to enable the Sine shape sonification (Figure~\ref{fig:sonifications_s2}(b)). As Synth had performed best in Study 1, we decided to further refine it for Study 2. Qualitative feedback from the study indicated that Synth had led to frustration during rough tumor localization. This was caused by the absence of sound feedback whenever the probe was outside the tumor area. This led us to include a beat sound into Sine to indicate the probe's distance to the seed even outside the tumor area. Similar to Beep, the beat frequency increased as the probe approached the seed. The well-received discrete sonification of the tumor shape via a continuous synthesizer sound was kept in the Sine sonification. Three output parameters from Wekinator were mapped to \textcolor{red}{the} volume and beat frequency of the seed sound as well as \textcolor{red}{the} volume of the tumor sound.

\subsection{Procedure}
The two conditions (Beep and Sine) were presented to each participant in a random order. We again began the study session by administering the pre-task questionnaire, followed by the tutorial phase. During the testing phase, the volunteers were tasked to mark the tumor margin, including the seed location on the silicone breast model. Once eight tumors were located, the \textcolor{red}{participants filled out the} post-task questionnaire. The same process was repeated for the second condition.

\subsection{Data Analysis}
The same measures as in Study 1 were collected. Due to a shift in experiment setup from a 2D to a 3D surface, the location of the margin and seed were recorded by tracing the drawn markings with the tracked probe. These locations were stored in the Unity coordinate frame and later processed in Matlab.
To obtain the ground truth (GT), we bottom-up projected the virtual 3D tumors located inside the virtual breast onto the breast model's surface in Unity. We stored this projection for each of the 15 different tumor objects. 
To compute the Sørensen–Dice coefficient, area ratio, and intercentroid distance in Matlab, the 3D GT and 3D marking recording (Figure~\ref{fig:matlab_plots}) were projected onto a 2D plane that best fits the 3D shape. The plane is calculated by minimizing the sum of the quadratic distances between the plane and the points constituting the 3D shape, and the fit is performed by computing the eigenvalues and vectors associated with the distribution of points.

\begin{figure*} [h!]
    \centering
    \subfloat[Sørensen-Dice coefficient (0-1; larger better)]
    {\resizebox*{5.7cm}{!}
    {\includegraphics{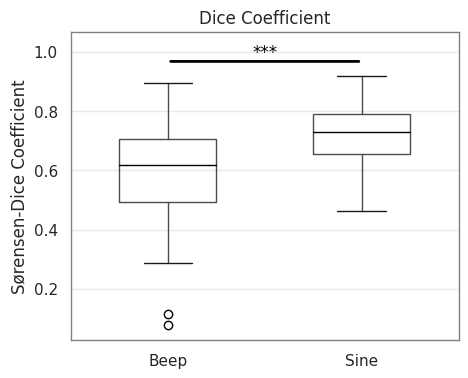}}}\hspace{6pt}
    \subfloat[Area ratio (1 equals perfect alignment of ground truth and drawn shape)]
    {\resizebox*{5.7cm}{!}{\includegraphics{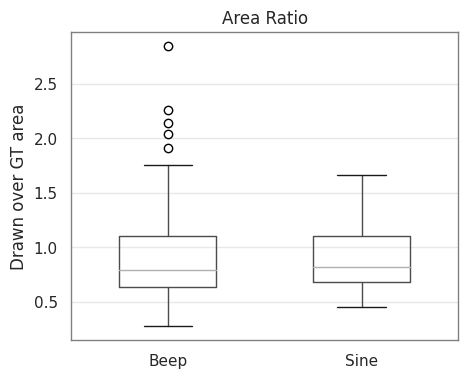}}} \hspace{6pt}
    \subfloat[Intercentroid distance (between ground truth and drawn seed location)]
    {\resizebox*{5.7cm}{!}{\includegraphics{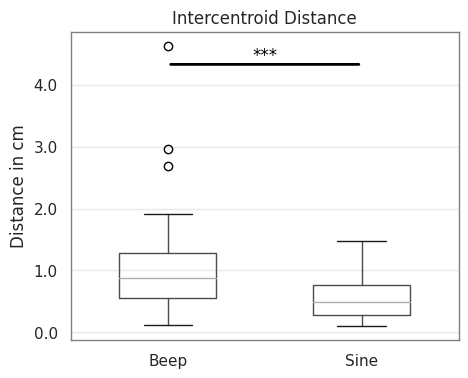}}}
    \caption{Box plots on localization accuracy in Study 2\textcolor{red}{. *** = p < 0.001}} 
    \label{fig:boxplots_2}
    \Description{Box plots of Sørensen-Dice coefficient (a), area ratio (b) and intercentroid distance (c) for the two sonification conditions Beep and Sine evaluated in Study 2.}
\end{figure*}

\subsection{Results}
\subsubsection{Accuracy}
A Shapiro-Wilk test showed a non-normal distribution for all accuracy measures in Study 2. \textcolor{red}{Eight o}utliers were removed using the interquartile range method. A Wilcoxon signed-rank test was used to determine significance. The shape sonification (Sine) achieved a significantly better (p $<$ 0.001) Dice coefficient (Figure~\ref{fig:boxplots_2}(a)) and significantly reduced (p $<$ 0.001) intercentroid distance (Figure~\ref{fig:boxplots_2}(b)) over Beep. The analysis showed no significant difference (p = 0.67) in area ratio between Beep and Sine (Figure~\ref{fig:boxplots_2}(c)). Thus, the shape sonification (Sine) was able to improve the accuracy of the localization task for both the tumor margin (Dice coefficient) and the seed location (intercentroid distance).

\subsubsection{Usability}
A Shapiro-Wilk test showed normality of the NASA-TLX data. Values outside the mean $\pm$ \textcolor{red}{three} std were \textcolor{red}{identified as outliers and one sample was excluded from the data}. A paired samples t-test showed no significant difference between sonifications for \textcolor{red}{the overall} NASA-TLX \textcolor{red}{score} (p = 0.08). \textcolor{red}{Out of the NASA-TLX subscales, only 'Effort' showed a significant difference. Beep resulted in significantly less (p $<$ 0.05) effort than Sine.} 
The SUS results were not normally distributed. 
\textcolor{red}{Zero o}utliers were \textcolor{red}{identified} using the interquartile range method. A Wilcoxon signed-rank test did not show significance. 
The accuracy and usability measures' means and standard deviations are reported in Table~\ref{tab:means_s2}.

\begin{table} [b]
    \caption{Means and standard deviations (std) of 
localization accuracy and usability for both sonifications in Study 2: Sørensen–Dice coefficient ([0,1]; larger better), area ratio (closer to 1 better), intercentroid distance (in cm; smaller better), \textcolor{red}{overall} task load\textcolor{red}{, individual NASA-TLX subscales }([0, 100]; smaller better) and usability (SUS, [0, 100]; larger better)}
    \label{tab:means_s2}
    \begin{adjustbox}{max width=\linewidth}
    \begin{tabular}{lcc}
        \toprule
        n = 21 & Beep & Sine \\
        \midrule
        Sørensen–Dice Coefficient & 0.60 $\pm$ 0.16 & 0.72 $\pm$ 0.10 \\
        Area Ratio & 0.92 $\pm$ 0.44 & 0.90 $\pm$ 0.29 \\
        Inter-centroid Distance (cm) & 0.99 $\pm$ 0.62 & 0.56 $\pm$ 0.34 \\
        NASA-TLX \textcolor{red}{- Overall} & 36.92 $\pm$ 13.67 & 44.35 $\pm$ 12.68 \\
        \textcolor{red}{NASA-TLX - Mental Demand & \textcolor{red}{37.86 $\pm$ 5.55} & \textcolor{red}{51.00 $\pm$ 4.57}}\\
        \textcolor{red}{NASA-TLX - Physical Demand & \textcolor{red}{23.57 $\pm$ 3.12} & \textcolor{red}{28.50 $\pm$ 4.57}} \\
        \textcolor{red}{NASA-TLX - Temporal Demand & \textcolor{red}{26.67 $\pm$ 4.56} & \textcolor{red}{28.42 $\pm$ 3.79}} \\
        \textcolor{red}{NASA-TLX - Performance & \textcolor{red}{60.25 $\pm$ 5.71} & \textcolor{red}{71.05 $\pm$ 4.74}}\\
        \textcolor{red}{NASA-TLX - Effort & \textcolor{red}{43.57 $\pm$ 5.01} & \textcolor{red}{58.95 $\pm$ 3.78}} \\
        \textcolor{red}{NASA-TLX - Frustration & \textcolor{red}{32.38 $\pm$ 6.46} & \textcolor{red}{30.50 $\pm$ 4.92}} \\
        System Usability Scale & 76.25 $\pm$ 16.23 & 77.50 $\pm$ 20.05 \\
        \bottomrule
    \end{tabular}
    \end{adjustbox}
\end{table}

\subsubsection{Qualitative Feedback}
The reported lower system usability of \textcolor{red}{Beep} was also reflected in the qualitative feedback from the post-study questionnaire. Participants enjoyed that Beep was \textit{"intuitive and easy to use"}, \textit{"much faster"}, and \textit{"not mentally challenging"}. However, they disliked that no information about the size and shape of the tumor was provided. One participant wrote that Beep \textit{"lacks precision and information"}. The person further mentioned that this results in a \textit{"lack of confidence using it"}. Another participant said: \textit{"I was not confident about my drawings, especially the size and boundaries of the tumor."}. Frustration was also expressed by another volunteer, who stated: \textit{"Discriminating between when I was near the seed vs when I was actually on it was rather taxing. It was also taxing to mentally translate the margin area through visuals alone."}. When asked how they would improve the Beep sound feedback, a few people suggested distinguishing between being near the seed versus being directly on top of it.

\textcolor{red}{86\% of the participants indicated to prefer the shape sonification (Sine) over Beep.}
The participants positively mentioned the \textcolor{red}{Sine} sonification's pleasantness of the sound, the accuracy gained through the information on the tumor margin, and the distinct nature of the two sounds. They reported to like the \textit{"clear separation along the border"}, the \textit{"continuous positive feedback if within the border"}, and that the \textit{"seed detection is easily distinguishable from the closeness sound"}. One participant also mentioned: \textit{"It’s very fun to use and offers more accuracy in identifying the tumor boundaries and seed."}.
The participants negatively pointed out the system's latency which caused a slight delay in the uptake of the sound. They further missed directional information: \textit{"I did not like that I got no directional cues - am I bottom left? am I top right?"}. When asked about ways to improve Sine, two participants mentioned making the tumor margin even more distinguishable: \textit{"The sound of the outline needs to be more obvious."}, \textit{"I would [...] make the device more sensitive to the outline of the tumor."}.

\subsection{Discussion}
To summarize, the Sine shape sonification resulted in significantly improved localization accuracy over Beep. \textcolor{red}{The task load data showed that Beep was significantly less effort to use than Sine. Although Sine received a higher system usability rating,} no significant difference in usability was found. However, given that shape sonification introduces another parameter and thus adds complexity to the Sine condition, \textcolor{red}{ an increase in effort} and a slight reduction in usability is not surprising. Furthermore, the use of shape sonification due to careful construction of the outline based on auditory feedback inevitably increases task time. One should investigate whether the sonification design or the prolonged task time is the \textcolor{red}{primary} driver of reduced usability in this scenario. Eliminating the slight latency present in the technical setup should additionally improve the system's usability.

Although the breast model enabled a replication of the real use scenario, the silicone was not soft enough to provide realistic breast tissue properties. A study interested in yielding clinically applicable results should look into alternative breast models. 
While this study included two volunteers with medical expertise, no breast surgeons were present. To gain insights \textcolor{red}{into} the sonifications' perception among the intended users of the system, we conducted a third and final study with breast surgeons.

%%%%%%%%%%%%%%% 3 3 3 3 3 3 3 3 3 3 3 3 3 3 %%%%%%%%%%%%%%%%%%%%

\section{Study 3 - 3D Surface with Breast Surgeons}

To further inform the development of the new auditory display and to test the system's suitability for the surgical task, we evaluate the refined shape sonification with the intended users of the system – the breast surgeons. Contrary to our prior studies, we also determined the shape sonification's influence on the accuracy of tumor excision. 

\subsection{Participants}
A total of four attending breast surgeons with a mean age of 44 $\pm$ 9 years std participated in the study, all four of which were women. None of them had a hearing impairment. \textcolor{red}{All four surgeons indicated extensive experience using the Savi Scout® seed based localization system during lumpectomy.}

\subsection{Apparatus}
Except for a change in the breast model, the hardware and software setup was identical to the prior studies (Figure~\ref{fig:3_studies}(c)).
In this study, the breast phantoms were comprised of agar, glycerol, and red food coloring. Agar and glycerol were selected for their replicability and compatibility with various imaging modalities such as ultrasound and MRI \cite{oglat}. The fabrication of the 16 agar phantoms consisted of continuously stirring a mixture of 3\% agar, 9.5\% glycerol, and 87.5\% distilled water on a hot plate for 8 minutes while ensuring that the mixture did not boil. Drops of red food coloring were added to make the tumors indistinguishable from the breast model mass. 16 tumors were poured from the same mixture as the breast model. However, a 1:200 diluted MRI contrast agent (Feraheme) was added to the mixture for visibility of the tumor mass on MRI imaging. The tumors were inserted into the breast pour after a curing time of about one hour. Diverse tumor shapes and sizes, as well as placements inside the breast, were chosen for a realistic scenario. The phantoms were wrapped in cling wrap and stored in a refrigerator to prevent the agar from drying.

Before and after the study, a General Electric MRI scanner was used to scan all 16 agar models. The pre-study scans were then utilized to segment and save the breast and tumor volumes as OBJ files. These virtual objects were subsequently imported into Unity to align the real and virtual experimental setups. This enabled us to determine the probe's position in relation to the breast and tumor.

\begin{figure*} [h]
    \centering
    \includegraphics[width=\textwidth]{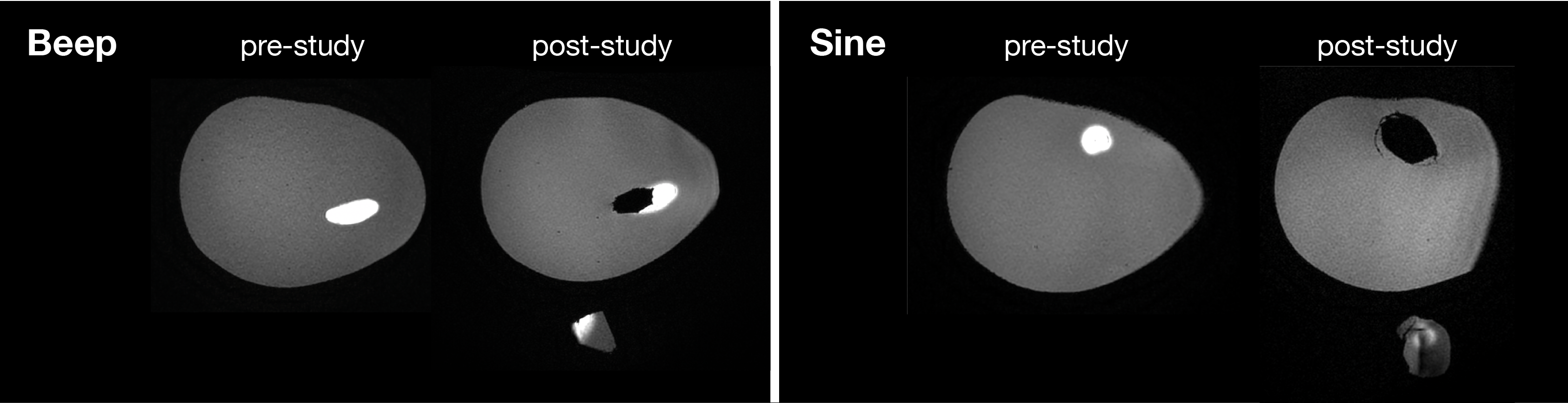}
    \caption{Pre- and post-study MRI images of two exemplary agar breast phantoms. The white areas represent the tumor mass. \textcolor{red}{In} the post-study images, the resected volumes are shown below the breasts. Tumor tissue remains in the breast, where Beep was used. Complete tumor resection was achieved using Sine.} 
    \label{fig:mri}
    \Description{Beep was used for localization on the left breast, and Sine was used on the right breast.}
\end{figure*}

\subsection{Sonification}
The breast surgeons were presented \textcolor{red}{with} the same sonifications as the participants in Study 2:  Beep (Figure~\ref{fig:sonifications_s2}(a)), a replica of the clinical sound feedback, and the proposed shape sonification Sine (Figure~\ref{fig:sonifications_s2}(b)).

\subsection{Procedure}
Each surgeon filled out the pre-task questionnaire followed by an explanation and initial training step using the first sonification. Then, the surgeon was asked to first mark the tumor margin and seed location on the surface of the agar breast model, transfer the marking from the cling wrap onto the agar using a small spatula, and lastly, excise the tumor volume from the breast model using a scalpel and a small spoon-like tool. After they marked and removed the tumors from two breast models, they filled out the post-task questionnaire for the sonification that they had just used. The same process was repeated for the second sonification on two more breast models.

\subsection{Data Analysis}
The pre- and post-study MRI scans were used to determine the calculated resection ratio (CRR), a measure used in clinical research to define the amount of excess healthy breast tissue resected during lumpectomy \cite{bib13, bib14, shin}. CRR is calculated by dividing the total resection volume (TRV) over the optimal resection volume (ORV). ORV was defined as the tumor volume, including a 2mm safety margin of healthy tissue. TRV and ORV were obtained by segmenting the 3D tumor and excision volumes from the MRI scans using Horos\footnote{Horos (https://horosproject.org/)}, an open-source medical image viewer.

All other quantitative (Dice coefficient, area ratio, intercentroid distance) and qualitative measures (pre- and post-study questionnaire) were recorded and analyzed as in Study 2.

\subsection{Results}
\subsubsection{Accuracy}
A Shapiro-Wilk test showed a non-normal distribution for all accuracy measures in Study 3. \textcolor{red}{Zero o}utliers were \textcolor{red}{identified} using the interquartile range method. A Wilcoxon signed-rank test showed a significantly improved Dice coefficient (p < 0.001) of Sine over Beep. No significant difference in area ratio, intercentroid distance, or calculated resection ratios was found. Figure~\ref{fig:mri} shows two exemplary post-study MRI scans for Beep and Sine.

\begin{table} [b]
    \caption{Means and standard deviations (std) of localization accuracy, resection ratios and usability for both sonifications in Study 3: Sørensen–Dice coefficient ([0,1]; larger better), area ratio (closer to 1 better), intercentroid distance (in cm; smaller better), calculated resection ratios (CRR, in cubic centimeters (cm$^3$)); closer to 1 better), \textcolor{red}{overall} task load\textcolor{red}{, individual NASA-TLX subscales }([0, 100]; smaller better) and usability (SUS, [0, 100]; larger better)}
    \label{tab:means_s3}
    \begin{adjustbox}{max width=\linewidth}
    \begin{tabular}{lcc}
        \toprule
        n = 4 & Beep & Sine \\
        \midrule
        Sørensen–Dice Coefficient & 0.51 $\pm$ 0.20 & 0.74 $\pm$ 0.12 \\
        Area Ratio & 1.57 $\pm$ 0.82 & 1.17 $\pm$ 0.73 \\
        Inter-centroid Distance (cm) & 0.83 $\pm$ 0.44 & 0.58 $\pm$ 0.29 \\
        CRR (cm$^3$) (TRV/ORV) & 2.54 $\pm$ 1.10 & 2.43 $\pm$ 1.58 \\
        NASA-TLX \textcolor{red}{- Overall} & 29.72 $\pm$ 16.98 & 30.83 $\pm$ 18.94 \\
        \textcolor{red}{NASA-TLX - Mental Demand & \textcolor{red}{42.50 $\pm$ 2.38} & \textcolor{red}{38.75 $\pm$ 3.86}}\\
        \textcolor{red}{NASA-TLX - Physical Demand & \textcolor{red}{10.00 $\pm$ 0.82} & \textcolor{red}{8.75 $\pm$ 0.50}} \\
        \textcolor{red}{NASA-TLX - Temporal Demand & \textcolor{red}{11.25 $\pm$ 0.50} & \textcolor{red}{20.00 $\pm$ 3.92}} \\
        \textcolor{red}{NASA-TLX - Performance & \textcolor{red}{45.00 $\pm$ 5.72} & \textcolor{red}{63.33 $\pm$ 2.89}}\\
        \textcolor{red}{NASA-TLX - Effort & \textcolor{red}{23.33 $\pm$ 2.52} & \textcolor{red}{31.67 $\pm$ 3.21}} \\
        \textcolor{red}{NASA-TLX - Frustration & \textcolor{red}{46.25 $\pm$ 4.11} & \textcolor{red}{22.50 $\pm$ 4.20}} \\
        System Usability Scale & 65.63 $\pm$ 11.43 & 85.63 $\pm$ 10.68 \\
        \bottomrule
    \end{tabular}
    \end{adjustbox}
\end{table}

\subsubsection{Usability}
A Shapiro-Wilk test showed a non-normal distribution of the NASA-TLX and SUS data. 
\textcolor{red}{T}he interquartile range method \textcolor{red}{was used to identify outliers, and one sample was removed from the data}. Due to the small sample size of four surgeons, the  Wilcoxon signed-rank test showed no significant difference between the two sonifications for neither task load nor system usability. However, the shape sonification Sine achieved a better usability rating than Beep. The accuracy and usability measures' means and standard deviations are reported in Table~\ref{tab:means_s3}.

\subsubsection{Qualitative Feedback}
This improvement was also echoed by the qualitative feedback we received in the post-task questionnaire. Their feedback on the sonification they routinely use in surgery (Beep) was that: \textit{"It only helps localize the seed, so I have to estimate and guess tumor size and shape."}.Another surgeon wrote: \textit{"It was harder to distinguish the edge of the lesion with as much confidence as when the Sine was on."}.
\newline
When asked to rank the two sonifications by preference, all surgeons reported \textcolor{red}{preferring} Sine. They liked that the \textit{"[tumor] mass was integrated into the sounds"} and that \textit{"the dual sounds"} gave them a \textit{"second way to verify that the target is within the specimen."} They further said about Sine: \textit{"It helps give shape and size approximation of the targeted tumor."}.

\subsection{Discussion}
A comparison of the results from the surgeon study to the non-surgeon study (Study 2) shows differences in usability results. While non-surgeons rated the task load of Sine (44.35$\pm$12.68) to be higher than Beep, the surgeons' task load rating of Sine (30.83$\pm$18.94) was much lower and rated as almost equally demanding as Beep (29.72$\pm$16.98).
Sine received a better usability rating (85.63$\pm$10.68) than Beep (65.63$\pm$11.43) by the surgeons. Although the small number of participants does not allow for a generalized claim, the trend in the reported data shows the great potential of shape sonification in improving the usability of the breast cancer localization task.

Overall, Study 3 was limited by the small sample size. A study with more breast surgeons should be performed to achieve significant results. A future study should also look into ways to improve the breast models' flexibility and thus increase the similarity with breast tissue. In addition, the wet surface of the agar model resulted in movement of the cling wrap. This, in turn, added imprecision to the accuracy results. An alternative method for surface markings should be considered.

\section{Discussion}

%H1 - increased localization accuracy
Our study results were able to verify \textbf{H1}. Shape sonification \textcolor{red}{increased} the accuracy of the breast tumor localization task. All three studies reported a significantly improved Dice coefficient, indicating a greater overlap of the ground truth (GT) and drawn tumor shapes. Study 1 additionally reported a significantly reduced area ratio while, Study 2 showed a significantly reduced distance between the GT and drawn seed location.

The increased localization accuracy suggests \textcolor{red}{shape sonification's potential to reduce the reoperation rate} due to tumor-positive margins. A comparable experiment on breast tumor localization using visual Augmented Reality guidance reports an average Dice coefficient of 0.2 \cite{perkins2}. Our proposed shape sonifications achieved an average Dice coefficient of 0.57, 0.72, and 0.74 in Studies 1, 2, and 3, respectively, proving superior accuracy.

%H2 - increased usability
We were furthermore able to partially confirm \textbf{H2}. Study 1 demonstrated significantly increased usability of the Synth shape sonification over the current clinical sound feedback. The combination of two distinct sounds - a continuous and a beat sound - improved discrimination between margin and seed location. Studies 2 and 3 showed no significant differences in task load or usability between shape sonification and the status quo auditory feedback. This effect might be attributed to the added complexity and increased task time that comes with reconstructing a shape from sound feedback. 

Interestingly, the breast surgeons in Study 3 \textcolor{red}{reported improved} usability of the shape sonification (85.63$\pm$10.68) over Beep, the status quo sonification (65.63$\pm$11.43). Moreover, shape sonification was ranked as the preferred option in all three studies. This gives hope that improvements \textcolor{red}{to} the sonification design can enhance the usability of the breast cancer localization task and make sound-based guidance systems more user-friendly than current alternatives.

%H3 - reduced overexcision
We were not able to prove \textbf{H3}'s claim of reduced over-excision ratios when using shape sonification. Due to the limited number of breast surgeons at our University hospital and the copious resources needed to fabricate and obtain MRI scans of the breast models, our sample size was too small to achieve statistical significance. Although insignificant, our data shows a slight reduction in \textcolor{red}{the} removal of excess healthy breast tissue (2.43$\pm$1.58) over the current auditory feedback (2.54$\pm$1.10). The results for both sonifications fall within the clinically reported average ratios of 2.3 to 2.5 of total resection volume over ideal resection volume \cite{bib11, bib14}.
This non-significant reduction of the resection ratio can be attributed to the lack of depth feedback during excision. This might have led surgeons to take more volume than necessary to ensure resection of the entire tumor mass.

%\textcolor{red}{Two exemplary research studies have demonstrated that auditory feedback is effective in localizing breast cancer in both male and female patients \cite{tingen2020savi, gao2019breast}. Their findings indicate that auditory displays, including our proposed shape sonification model, are viable options for breast tumor localization in both male and female patients.}

\subsection{Design Implications}

The presented auditory display was designed with the intention of smooth integration into the surgical workflow of a lumpectomy. Instead of replacing current localization systems and forcing the surgical team to adapt to a new workflow, the \textcolor{red}{proposed} sonifications build on the current seed-based localization system, requiring only minor adjustments on the side of the breast surgeon. 
Due to the introduction of additional tumor margin sonification, the localization task will likely be prolonged. However, the consecutive surgical task of tumor excision might shorten, as gradual excision due to guessing the tumor margin location can be eliminated. 
These benefits of the presented auditory display over the traditional method will hopefully allow for fast and easy adoption into clinical routine.

% General learnings for HCI community
\textcolor{red}{While the introduced medical application led us to design an interactive shape sonification that prioritizes precise localization and reconstruction of a shape's contours, previous shape sonification works in the field of Human-computer interaction (HCI) have focused on 2D shape recognition \cite{sanchez, gerino}, 3D object recognition \cite{auvray, gholamalizadeh}, image understanding \cite{yoshida, meijer} or curvature perception \cite{alonso, boyer}. However, our evaluation has revealed general findings on shape sonification that contribute to the investigation of auditory displays in the field of HCI:}

\begin{enumerate}
\item\textcolor{red}{Shape sonification can provide sub-centimeter precision in localizing the contour of a 2D shape.}
\item \textcolor{red}{Encoding proximity to two objects in two sounds is preferred over one sound as it reduces the task load.}
\item \textcolor{red}{The use of distinguishable concurrent sounds (e.g. a beat and a continuous sound) can lead to improved usability of an auditory display for visuospatial tasks.}
\end{enumerate}

\textcolor{red}{
%Outside the introduced medical application, the shape sonification technique presented in this paper could be used for the design of sensory substitution devices. 
These findings are not only applicable to surgical sonification but are also of relevance to the design of sensory substitution devices. Like the works by Meijer [\citeyear{meijer}], Auvray et al. [\citeyear{auvray}], and Gerino et al. [\citeyear{gerino}], who present visual-to-auditory sensory substitution devices for the blind or visually impaired, our sonification technique could also provide visuospatial information to people with vision impairment. Sonified representations of a map, an image, or a user interface could be generated and explored using the presented interactive shape sonification. However, a user study with people with congenital blindness or visually impaired individuals who lost their eyesight later in life would be imperative to understand how blind people perceive and interact with the presented sonification techniques.
Besides the design of accessible technologies, our technique could be used to enrich multi-sensory interactions in Mixed Reality by creating more engaging sensory experiences. Shape sonification could, for example, enhance spatial understanding of virtual objects that are occluded or out of sight.}

\subsection{Limitations}

Current breast cancer localization systems do not incorporate tumor margin information \textcolor{red}{as only the location of the seed is known at the time of surgery.} 
\textcolor{red}{In our studies, we were able to bypass this issue by using breast models that allowed only slight deformation and by simulating the use of a radiological image acquired in the same position as the patient is in during surgery (e.g. supine MRI). However, this setup does not correspond to the real circumstances. To bridge the gap between experimental and patient application, a sophisticated tracking system is required. Advanced biomechanical modeling of breast tissue deformation enables these tracking systems for breast surgery. Modeling breast tissue deformation is a field of research on its own, which, for example, works by Samani et al. [\citeyear{samani}], Gavaghan et al. [\citeyear{gavaghan2008}], Eiben et al. [\citeyear{eiben2016}] or Alca{\~n}iz et al. [\citeyear{alcaniz2022}] are addressing. An integration of the valuable research results from this field and our work will be the key to a user-friendly and robust solution for breast cancer localization in clinical practice.}

%To know the real shape of the tumor at time of surgery, one would for example need to obtain a supine MRI scan (lying on the back, same position as surgery) or use soft tissue modeling to estimate the tumor shape and location. 

A limiting factor of our hardware and software setup is the use of three networks that stream data between the different software applications. This setup has caused a noticeable latency on the user's end and influenced the usability of the sonifications. Future studies might examine alternative technical setups for realizing the location to sound mapping while avoiding latency issues.

Another observed limitation of our study design is the user-dependent accuracy of the drawn markings. This accuracy was influenced by the individual strategies the participants adopted for marking. The need to remove the probe from the current position to mark the location in the center underneath the tip of the probe added to \textcolor{red}{the} imprecision of the drawn outlines. While this task design was consciously chosen as it represents the current clinical practice, it might be worth investigating how \textcolor{red}{the} marking of the tumor location can be improved. Methods to increase marking accuracy might include tracking the pen.

\subsection{Future Work}
Future work on an auditory display for breast cancer localization should, first of all, prioritize the integration of depth information into the design. Information on the tumor's exact position and depth within the located column of breast tissue is crucial in achieving the desired clinical outcomes. Including depth information might \textcolor{red}{significantly} impact the resection ratio and thus decrease the amount of excess healthy breast tissue unnecessarily removed today.

Secondly, the future design of such sonification models could benefit from including directional cues. These cues might make the localization and marking process faster and reduce user frustration. Yet this addition might increase the sonification's complexity and lead to \textcolor{red}{a} higher cognitive load.

Lastly, the issue of realistic simulation of breast tissue properties in an experimental setup should be tackled. This would include not only the use of breast models made from more flexible materials, but also a method for reliably registering and tracking the material's deformation over the course of the study.

\section{Conclusion}

The field of breast cancer localization is evolving, with surgical guidance systems showing great potential for improving patient outcomes. Compared to current localization systems that only provide sound feedback on the location of a marker implanted inside the tumor, our approach provides more comprehensive guidance through additional sonification of tumor margins.

In three user studies, we demonstrated that the proposed shape sonification significantly enhances localization accuracy. We were furthermore able to show that breast surgeons utilizing shape sonification experienced improved usability compared to the current clinical auditory feedback.

We showed evidence of the accuracy of auditory displays, \textcolor{red}{reinforcing} their potential as a substitution or addition to conventional, visual \textcolor{red}{user} interfaces. Our work presents an exemplary use case for shape sonification in a surgical precision task and hopefully helps to promote broader application of auditory displays and multimodal, e.g., audio-visual and audio-haptic, interfaces in surgical applications and beyond.

%\section{Acknowledgments}
\begin{acks}
We would like to thank the breast surgeons at Stanford Women's Cancer Center for their feedback and support throughout this project. We would further like to thank the Stanford Department of Neuroscience and the Department of Radiology for providing the workspace, materials and devices needed to make this project possible. 
\end{acks}

\bibliographystyle{ACM-Reference-Format}
\bibliography{bibliography}

%\appendix

\end{document}